\documentclass{article}
\usepackage{ijcai23}

\usepackage{times}
\usepackage{mathtools, amssymb, amsfonts}
\usepackage{soul}
\usepackage{multirow}
\usepackage{url}
\usepackage[hidelinks]{hyperref}
\usepackage[utf8]{inputenc}
\usepackage[small]{caption}
\usepackage{graphicx}
\usepackage{amsmath}
\usepackage{amsthm}
\usepackage{booktabs}
\usepackage{algorithm}
\usepackage{algorithmic}
\usepackage{xcolor}
\usepackage[switch]{lineno}
\usepackage{microtype}

\newcommand{\Rel}{\mathcal{R}}
\newcommand{\V}{V}
\newcommand{\Ein}{\ensuremath{E}}
\newcommand{\Epred}{\ensuremath{E_{\textrm{pred}}}}
\newcommand{\Etarg}{\ensuremath{E_{\textrm{target}}}}

\newcommand{\notimplies}{\;\not\!\!\!\implies}

\newcommand{\PotE}{\ensuremath{\Lambda}}

\title{
Musical Voice Separation as Link Prediction: \\ Modeling a Musical Perception Task as a Multi-Trajectory Tracking Problem
}

\author{
Emmanouil Karystinaios$^1$\textsuperscript{\textdagger}\and
Francesco Foscarin$^1$\textsuperscript{\textdagger} \and
Gerhard Widmer$^{1,2}$
\affiliations
$^1$Johannes Kepler University Linz, Austria\\
$^2$LIT AI Lab, Linz Institute of Technology, Austria\\
\textsuperscript{\textdagger} Equal contribution among authors.
\emails
\{firstname\}.\{lastname\}@jku.at
}

\begin{document}

\maketitle

\begin{abstract}

This paper targets the perceptual task of separating the different interacting voices, i.e., monophonic melodic streams, in a polyphonic musical piece. 
We target symbolic music, where notes are explicitly encoded, and model this task as a Multi-Trajectory Tracking (MTT) problem from discrete observations, i.e., notes in a pitch-time space.
Our approach builds a graph from a musical piece, by 
creating one node for every note, and separates the melodic trajectories by predicting a link between two notes if they are consecutive in the same voice/stream. 
This kind of local, greedy prediction is made possible by node embeddings created by a heterogeneous graph neural network that can capture inter- and intra-trajectory information. Furthermore, we propose a new regularization loss that encourages the output to respect the MTT premise of at most one incoming and one outgoing link for every node,
favouring monophonic (voice) trajectories; this loss function might also be useful in other general MTT scenarios.
Our approach does not use domain-specific heuristics, is scalable to longer sequences and a higher number of voices, and can handle complex cases such as voice inversions and overlaps.
We reach new state-of-the-art results for the voice separation task in classical music of different styles.~\footnote{All code, data, and pretrained models are available on \url{https://github.com/manoskary/vocsep_ijcai2023} }

\end{abstract}

\section{Introduction}


%

The Multi-Trajectory Tracking (MTT) problem considers an unknown number of moving objects and deals with the task of connecting a sequence of observations, usually points or short tracks in a spatiotemporal space, into accurate long-term trajectories.
MTT is a subject of study both in cognitive science and engineering areas and has applications in numerous fields, including guidance systems, surveillance, and threat assessment~\cite{van2016comparative}.

Existing approaches are based on dynamic programming algorithms that try to minimize the global cost (or maximize the global probability) of assigning observations to a certain trajectory~\cite{han2004algorithm,chong2009efficient,castnnon2011multi,van2016comparative}. 
Approaches based on deep learning have been developed in the related field of Multi-Object Tracking (MOT), which also concerns itself with an object identification step, usually from images or similar data where the object positions are not explicitly encoded. 
Together with the object detection modules, an MOT system also contains a tracking module, that needs to deal with the MTT problems. However, while MTT systems can rely only on the trajectory shapes~\cite{shooner2010high}, MOT systems can also rely on the similarity between the features extracted from the instantaneous state of the objects to compute the trajectory. For example, an MOT system that tracks a red car from video data will extract some features about the car being red that will greatly help distinguish this car from cars of other colours across frames.\footnote{A plethora of different terms, including ``Multi Target Tracklet Stitching'', ``Multi Target Tracking'', are used in the literature to identify different flavours of related problems, and their usage is not always consistent across different communities. In this paper, we will use only the two terms MTT and MOT with the meaning indicated above.}

\begin{figure}[t]
    \centering
    \includegraphics[width=0.99\columnwidth]{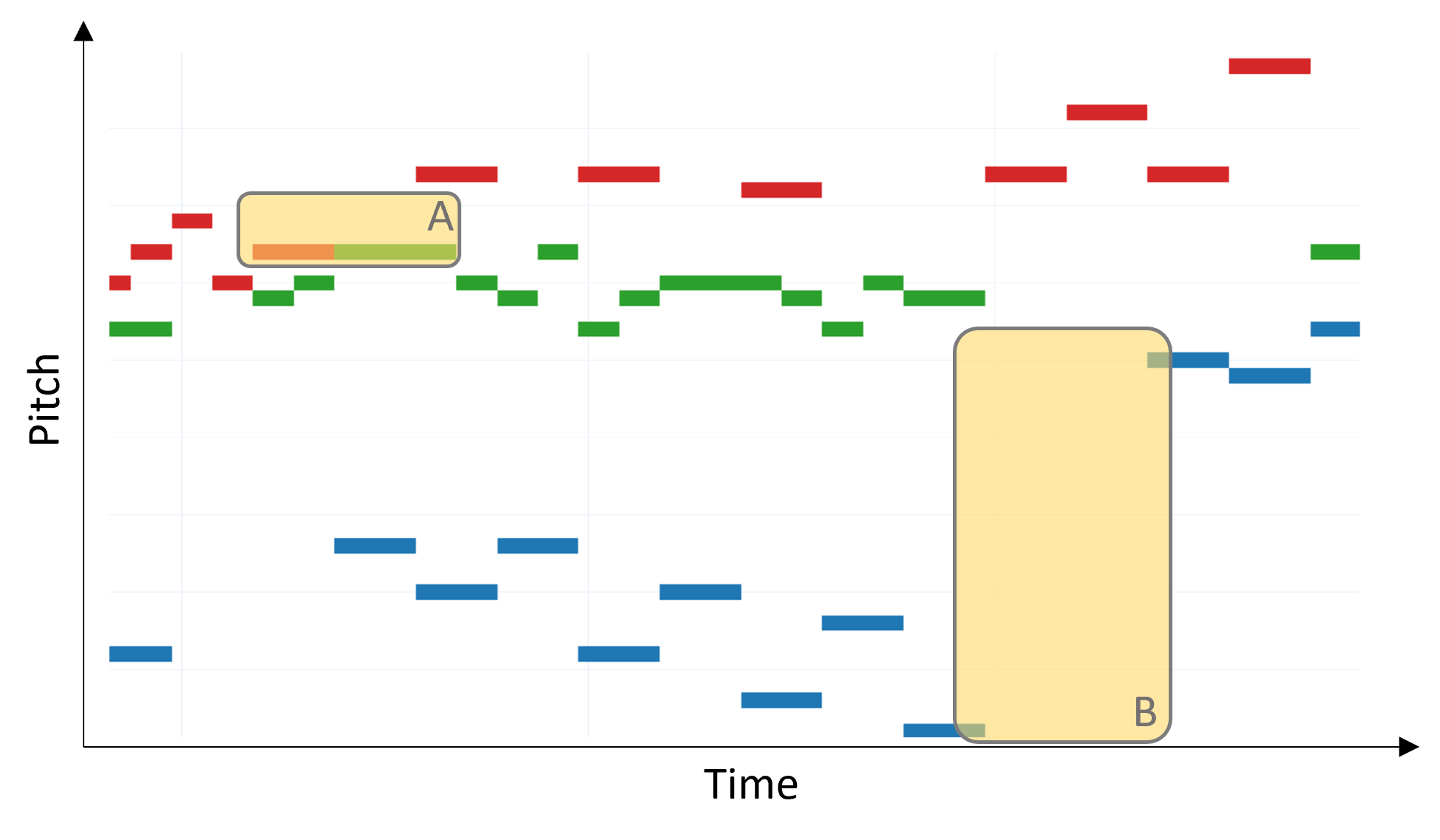}
    \caption{Example of multi trajectory following for musical voice separation in a pitch-time space.  Different trajectories are highlighted with different colors. Box (A) contains an example of consecutive notes with the same pitch belonging to different voices. Box (B) contains an example of ``distant'' notes belonging to the same voice. The musical excerpt is taken from Bach's Fugue in C-sharp major, BWV 872, measures 2-3-4.}
    \label{fig:MTT_example}
\end{figure}

In this paper, we will use an MTT approach to model a musical perception task,
namely, \textit{voice separation}. 
Many kinds of music can be seen as a sum or configuration of different \textit{voices}~\cite{aldwell2018harmony}, i.e., trajectories of (mostly) nonoverlapping notes (see Figure~\ref{fig:MTT_example}) as if they were produced by different voices singing together. Such voices are often not explicitly separated in a score (or a performance on a polyphonic instrument such as the piano), and the task of separating them is a useful step for a number of applications in music information retrieval, such as melody identification~\cite{ma2022robust} and MIDI to score transcription~\cite{foscarin2020musical}.

\textit{Symbolic music} denotes a set of musical formats (of different degrees of specificity and detail, e.g., MusicXML, **kern, MIDI), which contain explicit information about note pitches, onsets (i.e., starting time), and offsets (i.e., ending times)~\cite{foscarin2022match}, as opposed to audio files where all this information is mixed in a single acoustic signal.\footnote{In particular the input to our model is a set of notes with only pitch, onset, and offset (or duration) information. We assume them to be \textit{quantized}, i.e., notes whose onset and offset are aligned to a regular grid. It is irrelevant to our approach whether these notes are obtained from a score, a transcribed performance, a generation algorithm, or other sources.
}
Single notes can therefore be treated as trajectory observations in the pitch-time space, and voice separation can be framed as a trajectory following the problem. However, unlike the trajectories of objects that move in space, disentangling voice trajectories presents a set of unique challenges. Voices are not bound to stay in the immediate proximity of their last position and typically contain musical rests, that create ``holes'' in the trajectory.
Note that following the MTT and MOT differences we highlighted above, voice separation cannot be considered an MOT task. The static representative of a voice, i.e., a single note, does not contain any information about the voice to which it belongs. For example, two consecutive notes with the same pitch may belong to different voices (see Figure~\ref{fig:MTT_example}). The correct trajectory assignment can be made only by considering the rest of the trajectories.

Existing voice-following algorithms~\cite{chew2004separating,madsen2006separating,temperley2008probabilistic,duane2009streaming,jordanous2008voice,mcleod2016hmm} use perceptual principles and domain heuristics (with a few learned parameters in some cases) to compute the probability/cost of assigning a note to a certain voice, and then globally optimize the probabilities/costs for the entire piece.
These approaches have a set of intrinsic weaknesses that limit their performance. The first is a problem with generalization since the principles and heuristics employed may not remain valid for different kinds of music. The second is that music is a complex domain full of corner cases, with few rules, which would naturally fail in a number of situations. Moreover, the search space of dynamic programming approaches that do global optimization, with an unknown number of voices, scales exponentially with sequence length and the number of voices, making it necessary to work with short sequences or to add extra conditions to limit the search space (very common is to disallow voice crossings, though these can well occur in real music).


In this paper, we overcome these limitations with a technique based on \textit{Graph Neural Networks (GNNs)} that does not rely on any heuristic or domain principle. We frame the voice following as a link prediction problem; we model every note as a vertex in a graph and greedily predict a link between any pair of notes that should be consecutive in the same voice. This process results in a graph in which each fully connected group of nodes corresponds to a different voice.
To take advantage of inter- and intra-voice dependencies during the link prediction phase, we build a rich set of edges on top of the node vertices, based on the temporal relations between the corresponding notes in the music piece. 
We use edge relations to propagate local note information using heterogeneous message passing.
Since each link is independently predicted, our output could contain invalid configurations where a note has multiple incoming or outgoing links. Therefore, we also propose a new loss to enforce this number to be a maximum of one.
Our model consistently exceeds the state-of-the-art for Voice Separation on a large reference dataset with classical music of different styles. We can further increase the performance by running a polynomial-time global optimization algorithm that ensures that every note has a maximum of one incoming and one outgoing predicted link.

The contributions of this work are as follows.
\begin{itemize}
    \item a heterogeneous graph neural network approach to producing meaningful contextual features for the MTT task cast as a greedy link prediction problem;
    \item a new loss function to enforce the MTT constraint of a maximum of one incoming and one outgoing link for every trajectory observation.
    \item the application of MTT on symbolic music, resulting in a generalizable and scalable approach to the voice separation problem that handles overlaps and voice-switching;
    \item new state-of-the-art results on a reference dataset with classical music of different styles.
\end{itemize}

\section{Related work}
There are a number of approaches that address the problem of separating symbolic music into monophonic voices that are relevant to our work.
Duane and Pardo~\shortcite{duane2009streaming} propose the evaluation measure that was used in most of the subsequent research, including ours, and frame the voice separation problem as a set of link prediction problems between each pair of notes. The main conceptual difference between their approach and ours is that we enrich the note features with embeddings computed with a graph neural network, which drastically increases the quality of the predictions. Instead, they run a global optimization algorithm that scales exponentially with the note sequence length. This forces them to restrict the search space by not considering voice-crossings and to target only a few measures each time, stitching together the results afterwards.

The current state-of-the-art results, which we compare to in this paper, were produced by Mcleod and Steedman~\shortcite{mcleod2016hmm} with an HMM-based method where the probabilities of having a note assigned to a voice are based on Huron's~\shortcite{huron2001tone} perceptual principles of minimizing the time distance between consecutive notes and the pitch distance in a voice. To restrict the exponential search space for the global solution, they employ a modified Viterbi algorithm where at each step only the two best options are kept.

Notable for an approach which does not require the global optimization process, is the work of Gray and Brunescu~\shortcite{gray2016neural}. They run a left-to-right algorithm where a neural network greedily predicts, at each step, which existing voice a note should be assigned to (or whether to create a new voice). However, the network is not informed about inter-voice interactions or future voice trajectories, and the manually engineered features they use to help the prediction process are not enough to achieve higher experimental results than McLeod and Steedman.
Also worthy of mention is the work of Hsiao and Su~\cite{hsiao2021learning}, which model the score as a graph, and use unsupervised node graph clustering to separate different voices. The limits of this model lie in the fact that the clustering algorithm expects a given number of voices as input, and the heuristic the authors devise to estimate this number assumes the number of voices to be constant during the piece, which is a hypothesis that we are not introducing. Moreover, despite some slightly misleading claims, this approach does not reach new state-of-the-art results. 
in the case of quantized music, which we assume to be the input of our system. 

Another field of research \cite{cambouropoulos2006voice,kilian2002voice,rafailidis2009musical} targets music that can contain chords (i.e., multiple simultaneous notes) in the same voice. This is a different task (see \cite{cambouropoulos2008voice} for a discussion of different types of voice separation problems) and is not the focus of our work. We are also not targeting the problem of voice separation from human performance data that is explored by McLeod and Steedman~\shortcite{mcleod2016hmm}.

Similarly to the above-mentioned work in voice separation, multi-trajectory tracking (MTT) research is based on dynamic programming algorithms that perform global optimization on possible trajectories~\cite{han2004algorithm,chong2009efficient,castnnon2011multi,van2016comparative}. 
The field of Multi-Object Tracking (MOT) has received more attention in recent years, with a number of articles using GNNs.
Our approach shares some similarities with the work of Brasó and Leal-Taixé~\shortcite{braso2020learning}, Weng et al.\shortcite{weng2021ptp}, and Wang et al.\shortcite{wang2021joint}, in particular, the formulation of trajectory tracking as a greedy link prediction problem and the use of GNNs to generate relevant features for this prediction.
However, our data present a different set of challenges: the absence of useful static features and large temporal and spatial (pitch, in our case) gaps between consecutive observations in the same trajectory. Therefore, while the aforementioned MOT papers use only homogenous graphs (with some minor improvements by Brasó and Leal-Taixé~\shortcite{braso2020learning} that treat past and future links differently), we use heterogeneous graph neural networks with seven different link types, to create more informative node embeddings.

Finally, some works~\cite{weng2021ptp,wang2021joint} use the cross-entropy loss applied column-wise and row-wise in the adjacency matrix, to force each node to have a maximum of one incoming and one outgoing link. With the same goal, we propose an alternative loss that gives us better experimental results.

\section{Approach}
\begin{figure}
    \centering
    \includegraphics[width=\columnwidth]{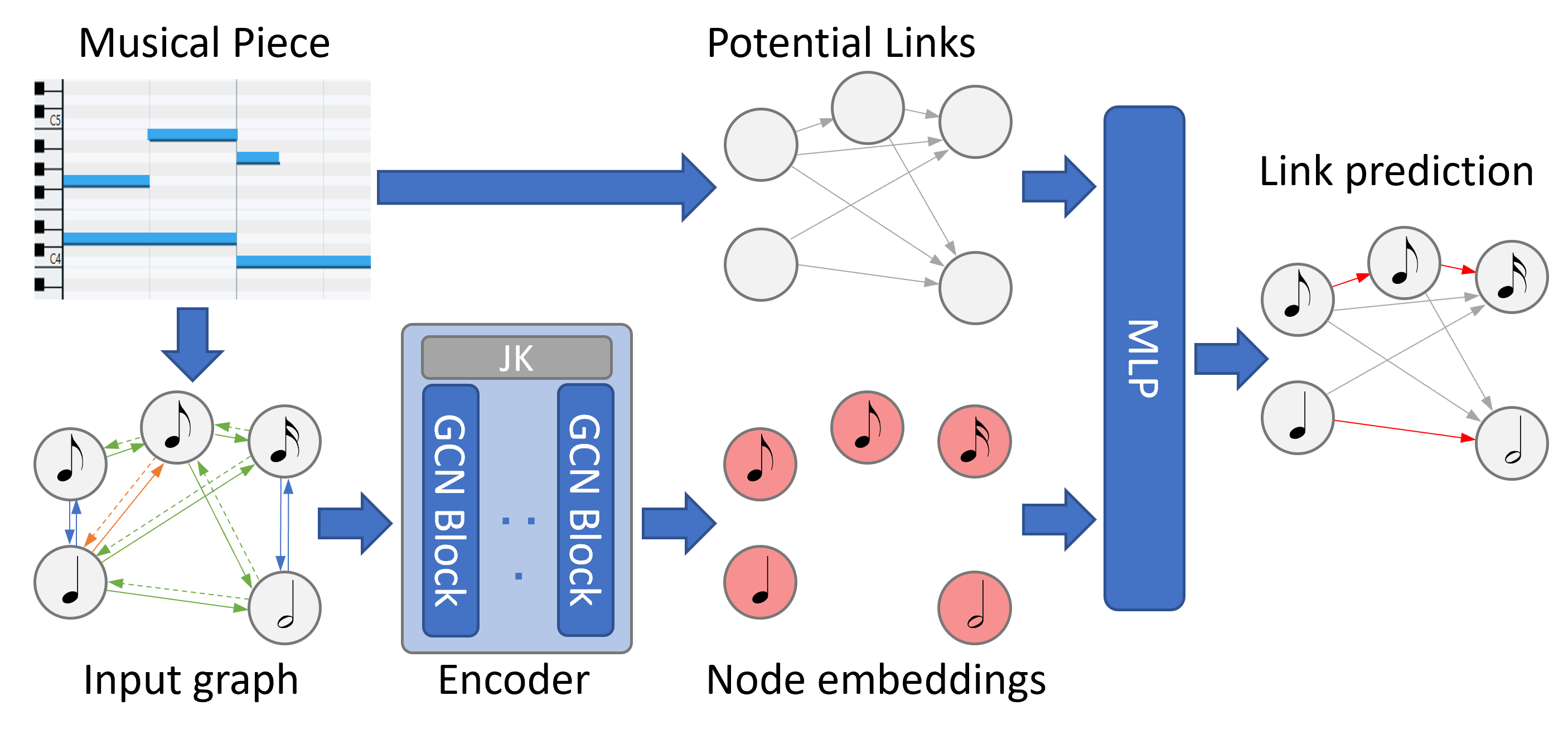}
    \caption{The GMTT model architecture for link prediction.}
    \label{fig:model}
\end{figure}

We model the input of our system, i.e., a set of quantized notes, 
with pitch, onset, and offset information, 
with a graph structure, where every note corresponds to a node in the graph. 
If we consider a set of links that connect only consecutive notes in the same voice (see Figure~\ref{fig:model}, right part), then the voices correspond to connected components in the graph.
Such a set of links is the desired output of our system, and the ground truth used for training.

Formally, consider a musical piece as a set $\V$ of notes, where the temporal position of the onsets defines a non-strict total order (i.e., multiple notes may have the same onset). Let $\Theta$ be a partition of $V$ in disjointed voices $\theta$. Since voices are monophonic, each voice defines a trajectory of length $l$, $\mathcal{T}_\theta = [ v^\theta_1 \dots v^{\theta}_l  \mid v \in V ]$: a strictly-ordered set that contains notes consecutive in the same voice. We can also view it as a set of pairs of consecutive notes in the same voice:

\begin{equation}
    E_{\textrm{target}} = \{ (v^\theta_{i}, v^\theta_{i+1}) \mid \forall \theta \in \Theta, \; v^\theta_{i} \in \mathcal{T}_\theta \}
    \label{eq:traj}
\end{equation}

$\Etarg$, as a specification of the voices in a piece, defines our ground truth.
Our goal is to predict such a set, and we do this by applying a binary classifier to every potential note pair $(u,v) \in V \times V$. We name the predicted set $\Epred$.
This process can be seen as predicting whether there is a link between $u$ and $v$,
hence it is usually called \textit{link prediction}.



We model a piece as a heterogeneous graph~\cite{hamilton2017representation} with different types of relations (or edge types) between notes and learn note embeddings using a GNN. Let $G = (\V, \Ein, \mathcal{R})$ be a graph such that $\V$ is the set of notes, $\Ein$ 
is the set of edge relations (or typed edges)
, and $\Rel$ is a set of relation types.
Every edge relation is defined by a triplet, i.e. $(u, r, v)$ such as $u, v \in \V$ and $r \in \Rel$. 
In addition, we associate every note with a vector of $k$ features that describe some intrinsic note properties. We assume all these feature vectors to be collected in a matrix $X \in \mathbb{R}^{|\V|\times k}$.

Therefore, our approach to voice separation can be summarized as follows: given a musical piece in symbolic form, we build a heterogeneous graph $G = (\V,\Ein,\Rel)$
and a set of node features $X$, and we use it to predict a set of links $\Epred$ that encode the voice trajectory according to Equation~\ref{eq:traj}.



\subsection{Graph Building}



Given a musical piece, the graph-building process creates a set of edges $\Ein$, with different relation types $\Rel$. 
We follow the work of Karystinaios and Widmer~\shortcite{cadence2022} and Jeong et al.~\shortcite{jeong2019graph} but adapt it to our voice separation problem by not using explicit musical rests (which are not present in our input) and considering a dedicated set of relation types 
that does not include any voice information.
Let us consider three functions $on(v)$, $dur(v)$, and $pitch(v)$ defined on a note $v\in V$ that extract the onset, duration, and pitch, respectively. 

A labeled edge $(u, r, v)$ of type $r$ 
between two notes $u, v$ belongs to $\Ein$ if the following conditions are met:
\begin{itemize}
    \item notes starting at the same time $on(u) = on(v) \xrightarrow{} r = \textrm{onset}$ 
    \item note starting while the other is sounding $on(u) > on(v) \land on(u) \leq on(v)+dur(v) \xrightarrow{} r = \textrm{during}$ 
    \item note starting when the other ends $on(u) + dur(u) = on(v) \xrightarrow{} r = \textrm{follow}$
    \item note starting after a time frame when no note is sounding $on(u) + dur(u) < on(v) \land \nexists v' \in V, \; on(v') < on(v) \land on(v') > on(u) + dur(u) \xrightarrow{} r = \textrm{silence}$
\end{itemize}


It is worth noting that, by construction, ``onset'' is the only undirected relation type, i.e., if $(u, \textrm{onset}, v)$, then $(v, \textrm{onset}, u)$. 
To keep our graph informed about the temporal evolution of the piece, we want the edges that connect notes at different times to be directed and only point to the future. But we also want the prediction to depend on future context. Therefore, we add in $\Rel$ three more types corresponding to the inverse of ``during'', ``follow'', and ``silence'', and we add to $\Ein$ the inverse edges with such types.


\subsection{Node Features}
We build a feature vector for every note $v \in V$. This contains the note's pitch-class and octave, as one-hot vectors of size $12$ and $8$ respectively, and the duration.
The duration is encoded as a single float value $d \in [0,1]$ computed as 
\begin{equation}
    d_n = 1 - tanh\frac{dur(v)}{dur(m)}, 
\end{equation}
where $dur(m)$ is the duration of the bar to which the note belongs. This information is also available in the symbolic music that we have as input. Normalization with bar duration has the objective of making $d$ independent of the time signature of the piece, and the $tanh$ function gives more resolution for lower note values while still being able to encode high durations.
Additionally, we use a positional encoding based on the 20 first eigenvectors from the Laplacian of the adjacency matrix similar to the work of Dwivedi et al.~\shortcite{dwivedi2020benchmarking}.

\subsection{Model}

Our model consists of two parts: a node encoder and a link predictor. 

The goal of the node encoder is to project node features $X$ to an embedding space that is enriched with context information. It consists of a series of
Residual Gated Convolutions~\cite{bresson2018residual} with Jumping Knowledge~\cite{xu2018jumping}. To account for the different edge relation types $r \in \Rel$, we compute independent representations for each type of relation and average them at the end of each convolutional block~\cite{schlichtkrull2018modeling}. More specifically, for every $v \in V$:

\begin{align}
    \mathbf{h}_{v_r}^{(l+1)} &= \mathbf{W}_1^{(l)} \mathbf{h}^{(l)}_v +
\sum_{u \in \mathcal{N}_r(v)} \eta^{(l)}_{v,u} \odot \mathbf{W}^{(l)}_2 \mathbf{h}^{(l)}_u \\
    \eta^{(l)}_{v,u} &= \sigma(\mathbf{W}^{(l)}_3 \mathbf{h}^{(l)}_v + \mathbf{W}^{(l)}_4
\mathbf{h}^{(l)}_u) \\ 
    \mathbf{h}_{v}^{(l+1)} &= \frac{1}{|\Rel|} \sum_{r \in \Rel} \mathbf{h}_{v_r}^{(l+1)}
\end{align}
where $h_v^{(l)}$ is the embedding of node $v$ for layer $l$, $\sigma$ denotes the sigmoid function, and $r$ denotes the relation type. 
On top of our convolutional blocks, we add Jumping Knowledge,
i.e., a bi-directional LSTM that connects the output of every convolutional layer $l \in L$ of the encoder:

\begin{equation}
    \mathbf{h}^{(jk)}_v = \sum_{l=1}^L \alpha_v^{(l)} \mathbf{h}_v^{(l)}
\end{equation}
where $\alpha$ denotes the attentional weights obtained by the LSTM and $\mathbf{h}_v$ is the node embedding for node $v$.

The link predictor part of our model is a multilayer perceptron (MLP) that performs the binary classification task of deciding whether two notes should be linked (i.e., be part of $\Epred$).
Due to our problem definition, the predictor only considers the links between $(u,v)$ if $on(u) + dur(u) \leq on(v)$. We name the set of potential links $\Lambda$.
For every potential link 
$(u, v) \in \Lambda$ we concatenate the embeddings of $u$ and $v$ produced by the encoder and give them as input to the link predictor:
    \begin{equation}
        \hat{y}_{u, v} = \textrm{MLP}(\textrm{concatenate}(h^{(jk)}_u, h^{(jk)}_v))
    \end{equation}
where $\hat{y}_{u, v} \in [0,1]$ denotes the predicted probability of a link from $u$ to $v$. For links $(u,v) \notin \PotE$ we set $\hat{y}_{u, v} = 0$.
We choose to concatenate the encoder's embeddings instead of taking their 
product because our links are directed, i.e. $(u,v) \in \Epred \notimplies (v,u) \in \Epred$. 

After this process, we are left with a probability for each pair $(u,v)$ to be part of $\Epred$. 
We round the probabilities according to a threshold value $\tau$ to obtain hard predictions.


\subsection{Loss}\label{subsec:loss}

In the training phase, we use positive and negative link prediction samples by subsampling the negative samples in $\PotE$ (i.e. 
$\{a \in \PotE \mid a \notin \Etarg \}$) 
to match the number of positives. We train our model by minimizing the Binary Cross-entropy: 

\begin{equation}
    \mathcal{L}_{\textrm{clf}} = -\sum_{u,v\in \mathcal{D}}\left( y_{u,v}\log(\hat{y}_{u,v}) + (1-y_{u,v})\log(1-\hat{y}_{u,v})) \right)
\end{equation}
where $y_{u,v} = 1$ if $(u,v) \in \Etarg, 0$ otherwise.


Let $\hat{A} \in [0,1]^{|V| \times |V|}$ 
be the  weighted graph adjacency matrix over $V$ that contains $\hat{y}_{u,v}$ at the corresponding indices. 
Since all our voices are disjointed, a note $u$ can be connected to at most one note $t$ that occurs before $u$ and at most one note $v$ that occurs after $u$.
This means that, in a perfect prediction scenario, we would have in $\hat{A}$ 
 only one non-zero element for each row and each column (or all zeros if the corresponding note ends or starts a voice, respectively).
One can therefore regard the ideal output of our system as the result of a linear assignment problem~\cite{burkard1999linear} over the predicted adjacency matrix $\hat{A}$.

In order to drive $\hat{A}$ to be in this format we propose a regularization loss, loosely inspired by~\cite{liu2022glan}, in addition to the classification loss. 
It is defined as follows:

\begin{align}
    \mathcal{L}_{reg}^{(1)} &= \| \boldsymbol{\zeta} - \sum_{i \in N} \hat{A}[i, \; :] \|_2 + \| \boldsymbol{\xi} - \sum_{j \in N} \hat{A}[:,\; j] \|_2 \label{eq:la_loss} \\
    \mathcal{L}_{reg}^{(2)} &= \| \boldsymbol{\zeta} - \sqrt{\sum_{i \in N} \hat{A}^2[i,\; :]} \|_2 + \| \boldsymbol{\xi} - \sqrt{\sum_{j \in N} \hat{A}^2[:,\; j]} \|_2 \label{eq:sparse_loss}\\
    \mathcal{L}_{reg}^{(tot)} &= \frac{\mathcal{L}_{reg}^{(1)} + \mathcal{L}_{reg}^{(2)}}{N} \label{eq:reg_loss}
\end{align}
where $N$ is the number of nodes in the graph, $\boldsymbol{\zeta}$ is a binary-valued vector of length $N$ with ones only for the nodes that are source nodes of ground truth links in $\Etarg$, and $\boldsymbol{\xi}$ is also a binary-valued vector, with ones only on the destination node indices of the ground truth links.

Equation \ref{eq:la_loss} encodes the linear assignment optimization objective, modified to allow rows and columns with only zeros. Furthermore, we add Equation \ref{eq:sparse_loss}, which uses the $L2$ norm of rows and columns, by squaring the positive valued $\hat{A}$. $L2$ and $L1$ are known to have different strengths in minimization problems~\cite{hastie2009elements}, and we found their sum to yield the best experimental results.
Together, \ref{eq:la_loss} and \ref{eq:sparse_loss} constitute the regularization loss which is normalized by the order of the graph, i.e. the number of nodes $|\V|$, since different musical pieces have a different number of notes.
The total loss of the system is then defined as:

\begin{equation}
    \mathcal{L}_{total} = \mathcal{L}_{\textrm{clf}} + \alpha \mathcal{L}_{reg}^{(tot)}
\end{equation}

where $\alpha$ is the regularization loss weight. The regularization loss weight $\alpha$ is initialized to $0$ and then it is gradually increased every epoch. Conceptually, during the first epochs of training, the focus is on the classification loss, but as training progresses the focus shifts towards a matrix that also satisfies linear assignment conditions.

\subsection{Postprocessing}
 
Based on the premise introduced in the previous section, we can view the predicted adjacency matrix $\hat{A}$ as a weighted matrix on which we can apply the Hungarian algorithm~\cite{crouse2016implementing} to solve the linear assignment problem.

Due to our restriction on the potential links $\Lambda$, the lower triangular and the diagonal of the adjacency $\hat{A}$ only contain zeros and should not be the focus of the prediction nor the assignment. Therefore, the linear assignment of our matrix only takes part in the upper triangular part of the predicted adjacency. This formulation simplifies the time complexity of the linear assignment.

Given the number of nodes $N$, our linear assignment optimization objective is defined as:
\begin{align*}
    \text{maximize} & \sum_{i = 0}^{N-1} \sum_{j=i+1}^{N} \hat{A}[i, j] * B[i,j] \\
    \text{subject to} & \sum_{j \in [i+1..N]} B[i,j]=1 \text{ for } i \in [0..N-1], \\ \text{ and }& \sum_{i \in [0..j]} B[i,j]=1 \text{ for } j \in [1..N]
\end{align*}
where $B[i,j]\in \mathbb{B}^{N \times N}$ is a learned binary mask over $\hat{A}$. The updated matrix is given for any two indices $i, j$ by $\hat{A}'[i, j] = \hat{A}[i, j] * B[i,j]$.
This matrix contains new link probabilities and, equivalently to the approach without post-processing, we round them according to a threshold value $\tau$ to obtain hard predictions.

\section{Experiments}

\begin{table*}[tbp]
    \centering
    \begin{tabular}{ l || c c c | c c c | c c c  }
         & & \textbf{McLeod} & & & \textbf{GMTT} & & & \textbf{GMMT+LA} &\\
         \hline
         \textbf{Datasets} & P & R & F1 & P & R & F1 & P & R & F1\\
         \hline
         Inventions & 0.992 & 0.991 & 0.992 & 0.989 & 0.997 & 0.995 & 0.996 & 0.995 & \textbf{0.997} \\
         Sinfonias & 0.982 & 0.982 & 0.982 & 0.987 & 0.989 & 0.978 & 0.987 & 0.982 &\textbf{0.985} \\
         WTC I & 0.964 & 0.964 & 0.964 & 0.949 & 0.983 & 0.967 & 0.980 & 0.973 & \textbf{0.976} \\
         WTC II & 0.964 & 0.964 & 0.964 & 0.945 & 0.979 & 0.962 & 0.976 & 0.968 & \textbf{0.972} \\
         Haydn & 0.781 & 0.781 & 0.781 & 0.787 & 0.929 & 0.850 & 0.883 & 0.860 & \textbf{ 0.872} \\
         \hline
    \end{tabular}
    \caption{Main results comparing the State-of-the-art on Voice separation with our approach. P stands for Precision, R stands for Recall, and F1 for F1-score. All the presented metrics are binary (only for the positive class, i.e. links). ($+$LA) stands for linear assignment postprocessing.}
    \label{tab:results}
\end{table*}

Below, we describe the datasets and the experimental settings.
\subsection{Datasets and preprocessing}
For training and testing our system, we need sets of quantized notes, with pitch, onset, and offset information, with a ground-truth separation into voices, provided by musical experts. We obtain these from a curated collection of musical scores\footnote{Note that a score also contains many other musical and graphical elements, such as rests, slurs, and stem directions, that could help in the voice separation task; we discard these in our application.}
from different composers and styles. In particular, we use all the 474 pieces 
from the Symbolic Multitrack Contrapuntal Music Archive (MCMA) (see~\cite{aljanaki2021mcma} for a detailed list of the pieces contained), and 662 pieces from the KernScore project \nolinkurl{http://kern.humdrum.org/}, in particular from the 
Bach Chorales, the Haydn string quartets, and the Mozart string quartets, which (mostly) satisfy our assumption of monophonic voices.
With a total of 1136 pieces, our dataset constitutes the largest data set publicly available for the voice separation task in symbolic music. 

To evaluate our system on different degrees of piece complexity and a variety of musical styles, we run five separate experiments. For each experiment, we fix as a test set a subset of our total data and consider 90\% of the remaining pieces for training and 10\% for validation.
In particular, our five test sets consist of 15 Bach inventions, 15 Bach sinfonias, 12 Bach fugues from WTC I, 12 Bach Fugues from WTC II, and 210 Haydn string quartet movements.
This corresponds with the data used by McLeod and Steedman~\shortcite{mcleod2016hmm}.

The pieces are available in different file formats and we use the Python library Partitura~\cite{partitura_mec} to extract the list of notes for each voice.
We then preprocess them by removing possible extra notes 
that would violate the monophonic voice assumption (mostly final chords at the very end of the pieces). In particular, if more than one simultaneous note exists in a single voice, we remove all except the highest. This preprocessing operation removes 0.72\% of the total notes in our dataset and was done similarly by Mcleod and Steedman~\shortcite{mcleod2016hmm}. 

For practical reasons, during the preprocessing phase, we also produce the set of potential links $\PotE$ that our model should predict on. For each note $u \in V$, we restrict the potential links to notes $v$ that are at most 2 measures after.
We may thus have $\PotE \subset \Etarg$, in which case some links in the ground truth will be assigned a probability 0, independently from any other model choice.
This means that if a sufficiently long rest exists between two notes of a voice, this voice will be inevitably split into two voices. It is worth noting that this happens very rarely in our datasets. However, this restriction can be relaxed to a longer duration, or removed completely, at the cost of a higher training and inference time.

The graphs produced by our preprocessing phase contain a total of 867,226 nodes and ~5M input edges. The number of potential links is ~30M, with 863,277 true links. From the latter, 2,264 links are not considered due to the modelling restrictions aforementioned.

\subsection{Main Experiment}\label{subsec:results}
For our main experiment on the five test sets mentioned above, we use the AdamW optimizer, with learning rate 0.003 and weight decay 0.005. We fix the other parameters of our model to 3 convolutional layers, a convolutional embedding size of 128, i.e. $h_u^{(l)} \in \mathbb{R}^{128}$ for a node $u$ and layer $l$, and prediction threshold $\tau=0.5$. The link predictor (MLP) part of our model has the same hidden size and number of layers.

We evaluate our model GMTT  with and without post-processing (i.e., applying the Hungarian algorithm to filter the model's predictions) and compare it with the current state-of-the-art voice separation method of McLeod \& Steedman~\shortcite{mcleod2016hmm}. The results are given in terms of recall, precision, and F1-score, calculated between the predicted links $\Epred$ and ground truth $\Etarg$.

The results in Table~\ref{tab:results} show that GMTT without post-processing is roughly on par with the SOTA, except for the Haydn String Quartet test set, for which we achieve significantly better results. This is an important result since in contrast to all the other approaches in the literature, our system is able to produce high-quality results for the problem of voice separation by only performing local/greedy predictions on single links. These performances are the result of embeddings that were generated by a graph neural network, which provided rich contextual information to the link predictor.

Linear assignment post-processing slightly reduces the recall, but considerably increases the precision, finally producing a higher F1-score. 
This means that, while our system comes very close to respecting the constraints of having only one incoming edge and one outgoing link for every note, especially thanks to our proposed regularization loss (see Section~\ref{sec:ablation} for a discussion on this), it still predicts some invalid configurations.

Overall, our improvement over the previous state-of-the-art approach is particularly significant for the Haydn String Quartets collection, which is also the most complex collection to separate. In Section \ref{sec:dicussion} below, we conduct a qualitative analysis of an individual example and discuss the musical elements that make this collection so challenging.

\subsection{Ablation studies}\label{sec:ablation}
We perform several ablation studies to understand how our design choices impact the model performance. For each experiment, we change one element of our architecture; if the element is useful, we expect a reduction in F1-score. 
We fix the hyperparameters of our model to the ones in Section \ref{subsec:results}. A summary of the ablation studies can be seen in Table \ref{tab:ablation} and the single experiments are discussed below.

\begin{table}[htbp]
    \centering
    \begin{tabular}{  l c c }
         \textbf{Models } & \textbf{Haydn} &  \textbf{WTC II} \\
         \hline
         Homogeneous & $0.809\pm 0.012$ & $0.943\pm 0.007$ \\
         SageConv Block  & $0.828\pm 0.005$ & $0.944\pm 0.002$\\
         No regularization  &  $0.720\pm 0.021$ & $0.856 \pm 0.049 $\\
         Fixed regularization  & $0.652\pm 0.18$ & $0.942\pm 0.015$\\
         \textbf{GMMT} & $\mathbf{0.850} \pm 0.001$ & $\mathbf{0.962}\pm 0.001$\\
         \hline
    \end{tabular}
    \caption{Ablation experiments, all the scores presented are binary F1-scores without postprocessing (i.e., LA). \textit{Homogeneous} denotes homogeneous graph message passing, \textit{SageConv} denotes the GraphSage convolutional block, \textit{No regularization} means a regularization weight $\alpha=0$, \textit{Fixed Regularization} has $\alpha=1$, and \textit{GMMT} is the model from Table 1, for comparison.}
    \label{tab:ablation}
\end{table}
\paragraph{Effect of Heterophily}
We tested our model using heterogeneous and homogeneous graph convolutional blocks. Leveraging the heterogeneous relations of the score graph has consistently improved results on voice separation, as can be seen by comparing row 1 (\textit{Homogeneous}) in Table \ref{tab:ablation} to row 5 (GMTT).

\paragraph{Effect of Convolutional Block}
We experimented with two types of Convolutional Blocks, standard Graph Sage and Residual Gated Convolution. Residual Gated Convolution consistently outperforms Graph Sage convolution in our datasets 
(row 2 vs.~row 5).

\paragraph{Effect of Regularization Loss}
The regularization loss with gradual weighting tends to stabilize training, resulting in more consistent results. Instead, using $\alpha = 1$ we may obtain results close to the final GMMT model using epoch weighting, but the results vary across repetitions of the run (row 4 in Table \ref{tab:ablation}). Using no regularization results in a high number of false positive predictions, which eventually drops the overall F1-score (row 3).
This clearly testifies to the ability of the regularisation loss to enforce row-column voice link sparsity. Our preferred model (row 5) has the lowest standard deviation and the best average performance across runs.

\section{Discussion}\label{sec:dicussion}

In this section, we investigate what makes voice separation on string quartets so challenging and the reasons for our model's better performance.
Compared to our other musical corpora, string quartets have a much different orchestration, which may include a bigger instrument range, more frequent and longer rests,
accompaniment split to multiple voices (for example between the cello and the viola), octave doubling, and big jumps in pitch within one voice/instrument line. Some of these differences can be attributed to the fact that the individual voices in a string quartet are played by different instruments, which makes it easier for listeners to keep track of voices and gives the composer more freedom in composing a texture that is still comprehensible.

\begin{figure}[t]
    \centering
    \includegraphics[width=\columnwidth]{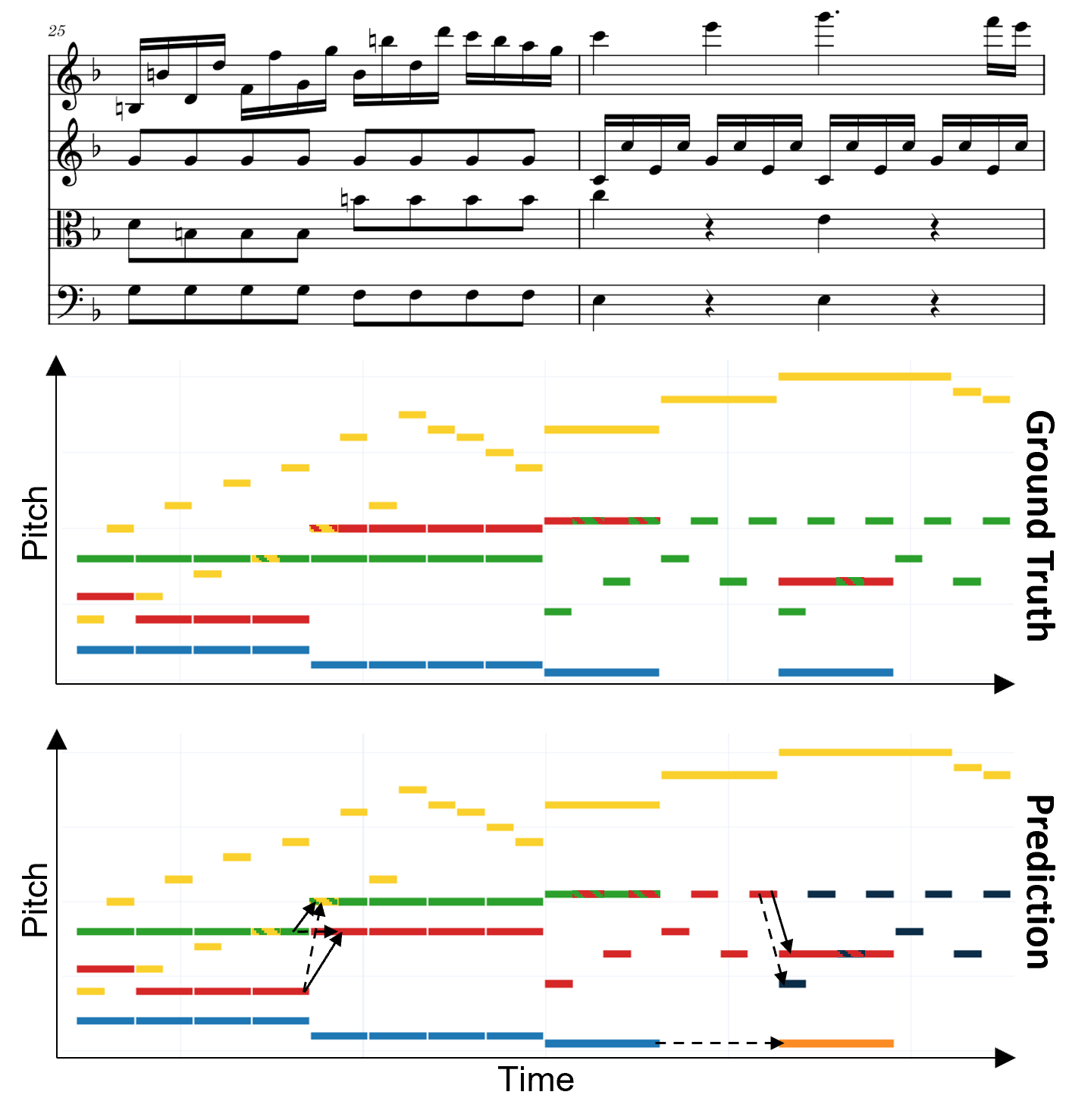}
    \caption{Ground truth and prediction for Haydn String Quartet Op17 No2, 1st mvt., bars 25-26. False negative errors are highlighted with dashed arrows and false positive with solid arrows. }
    \label{fig:quartet}
\end{figure}

In Figure \ref{fig:quartet}, we provide an example -- bars 25-26 of Haydn's String Quartet Op17 No2, 1st movement -- that demonstrates some of these properties together with our model's prediction and the ground truth. In this two-bar excerpt, we notice several challenging properties, such as octave jumps within the same voice, rests in between notes of the same voice, and voice crossings --
i.e., the 1st violin (top voice or yellow colour in the pianoroll) starts lower than the viola and the 2nd violin but ends more than one octave higher than any other voice by measure 26. This passage is a real challenge for the voice separation algorithm, and it is solved in a very reasonable way by our context-aware network.

A closer look at our model's errors reveals three faulty sections. The first occurs between the viola and the 2nd violin on the third quarter beat of bar 25, where we observe a rhythmically identical voice crossing (red and green streams). This is indeed hard to discern without additional knowledge; human listeners would likely parse this correctly based on the specific sound (timbre) of the instruments, performing something more akin to Multi-Object Tracking (MOT), in our terminology. Another voice crossing mistake occurs between the same two voices on beat 3 of bar 26, where we incorrectly predict the start of a new voice (which is then continued in black). Finally, we notice a rest discontinuity mistake on the cello (bottom; blue and orange lines) on the 3rd beat of bar 26. These examples test the limits of our network and also raise a perceptual question, whether a human analyst would be able to correctly assess these voice assignments from this bare representation, without additional perceptual clues. We leave this as an open question and a starting point for discussions on possibly more ``natural" definitions of the voice separation task.

\section{Conclusion and Future Work}

In this paper, we present a novel method for the perceptual problem of voice separation from symbolic music that achieves new state-of-the-art performance.
We propose a formulation as a multi-trajectory tracking problem, and an end-to-end approach, based on heterogeneous neural networks, that does not rely on any heuristic or musical assumption that may be correct only for limited kinds of music. That allows us to handle traditionally complex corner cases such as voice inversions and overlaps.
Furthermore, our approach can find a global voice separation solution on the entire piece, without pruning any potentially relevant option, with a complexity that is independent of the number of voices and scales polynomially with the number of notes in a piece. 
We also study the problem of reducing the dependence on any postprocessing algorithm and consider only the local, greedy predictions made by the neural network. We propose a new regularization loss that drastically improves our results in this setting, and may be useful for other general MTT scenarios. 

Our work can be extended in a number of directions. We plan to address voice separation for unquantized MIDI, i.e., musical pieces obtained by a human recording; this would let us explore how much the expressive timing and intensity deviations introduced by the performer correlate with voice information. Also, we want to relax the monophonic voice assumption to target pieces where multiple simultaneous notes (i.e., chords) can be present in the same voice. 
Finally, we aim at developing a voice separation method from raw audio, which could also contain relevant information for the voice separation task that is not available in the symbolic format. In the case of multiple instruments performing multiple voices, this would blend into the field of instrument/source separation.

\section{Acknowledgements}
This work was supported by the European Research Council (ERC) under the EU's Horizon 2020 research \& innovation programme, grant agreement No.\ 101019375 (\textit{Whither Music?}), and the Federal State of Upper Austria (LIT AI Lab).

\bibliographystyle{named}
\bibliography{ijcai23}

\begin{thebibliography}{}

\bibitem[\protect\citeauthoryear{Aldwell \bgroup \em et al.\egroup
  }{2018}]{aldwell2018harmony}
Edward Aldwell, Carl Schachter, and Allen Cadwallader.
\newblock {\em Harmony and voice leading}.
\newblock Cengage Learning, 2018.

\bibitem[\protect\citeauthoryear{Aljanaki \bgroup \em et al.\egroup
  }{2021}]{aljanaki2021mcma}
Anna Aljanaki, Stefano Kalonaris, Gianluca Micchi, and Eric Nichols.
\newblock Mcma: A symbolic multitrack contrapuntal music archive.
\newblock {\em Empirical Musicology Review}, 16(1):99--105, 2021.

\bibitem[\protect\citeauthoryear{Bras{\'o} and
  Leal-Taix{\'e}}{2020}]{braso2020learning}
Guillem Bras{\'o} and Laura Leal-Taix{\'e}.
\newblock Learning a neural solver for multiple object tracking.
\newblock In {\em Proceedings of the IEEE/CVF conference on computer vision and
  pattern recognition}, 2020.

\bibitem[\protect\citeauthoryear{Bresson and
  Laurent}{2018}]{bresson2018residual}
Xavier Bresson and Thomas Laurent.
\newblock An experimental study of neural networks for variable graphs.
\newblock In {\em Proceedings of the International Conference on Learning
  Representations}, 2018.

\bibitem[\protect\citeauthoryear{Burkard and Cela}{1999}]{burkard1999linear}
Rainer~E Burkard and Eranda Cela.
\newblock Linear assignment problems and extensions.
\newblock In {\em Handbook of combinatorial optimization}, pages 75--149.
  Springer, 1999.

\bibitem[\protect\citeauthoryear{Cambouropoulos}{2006}]{cambouropoulos2006voice}
Emilios Cambouropoulos.
\newblock Voice separation: theoretical, perceptual and computational
  perspectives.
\newblock In {\em Proceedings of the International Conference on Music
  Perception and Cognition (ICMPC)}. Citeseer, 2006.

\bibitem[\protect\citeauthoryear{Cambouropoulos}{2008}]{cambouropoulos2008voice}
Emilios Cambouropoulos.
\newblock Voice and stream: Perceptual and computational modeling of voice
  separation.
\newblock {\em Music Perception}, 26(1):75--94, 2008.

\bibitem[\protect\citeauthoryear{Cancino-Chac\'{o}n \bgroup \em et al.\egroup
  }{2022}]{partitura_mec}
Carlos~Eduardo Cancino-Chac\'{o}n, Silvan~David Peter, Emmanouil Karystinaios,
  Francesco Foscarin, Maarten Grachten, and Gerhard Widmer.
\newblock Partitura: A python package for symbolic music processing.
\newblock In {\em Proceedings of the Music Encoding Conference {(MEC)}},
  Halifax, Canada, 2022.

\bibitem[\protect\citeauthoryear{Castnn{\'o}n and
  Finn}{2011}]{castnnon2011multi}
Gregory Castnn{\'o}n and Lucas Finn.
\newblock Multi-target tracklet stitching through network flows.
\newblock In {\em Proceedings of the Aerospace Conference}. IEEE, 2011.

\bibitem[\protect\citeauthoryear{Chew and Wu}{2004}]{chew2004separating}
Elaine Chew and Xiaodan Wu.
\newblock Separating voices in polyphonic music: A contig mapping approach.
\newblock In {\em Proceedings of the International Symposium on Computer Music
  Modeling and Retrieval}. Springer, 2004.

\bibitem[\protect\citeauthoryear{Chong \bgroup \em et al.\egroup
  }{2009}]{chong2009efficient}
Chee-Yee Chong, Greg Castanon, Nathan Cooprider, Shozo Mori, Ravi Ravichandran,
  and Robert Macior.
\newblock Efficient multiple hypothesis tracking by track segment graph.
\newblock In {\em Proceedings of the International Conference on Information
  Fusion}. IEEE, 2009.

\bibitem[\protect\citeauthoryear{Crouse}{2016}]{crouse2016implementing}
David~F Crouse.
\newblock On implementing 2d rectangular assignment algorithms.
\newblock {\em IEEE Transactions on Aerospace and Electronic Systems},
  52(4):1679--1696, 2016.

\bibitem[\protect\citeauthoryear{Duane and Pardo}{2009}]{duane2009streaming}
Ben Duane and Bryan Pardo.
\newblock Streaming from midi using constraint satisfaction optimization and
  sequence alignment.
\newblock In {\em Proceedings of the International Computer Music Conference
  (ICMC)}, 2009.

\bibitem[\protect\citeauthoryear{Dwivedi \bgroup \em et al.\egroup
  }{2020}]{dwivedi2020benchmarking}
Vijay~Prakash Dwivedi, Chaitanya~K Joshi, Thomas Laurent, Yoshua Bengio, and
  Xavier Bresson.
\newblock Benchmarking graph neural networks.
\newblock {\em arXiv preprint arXiv:2003.00982}, 2020.

\bibitem[\protect\citeauthoryear{Foscarin \bgroup \em et al.\egroup
  }{2022}]{foscarin2022match}
Francesco Foscarin, Emmanouil Karystinaios, Silvan~David Peter, Carlos
  Cancino-Chac{\'o}n, Maarten Grachten, and Gerhard Widmer.
\newblock The match file format: Encoding alignments between scores and
  performances.
\newblock In {\em Proceedings of the Music Encoding Conference ({MEC})},
  Halifax, Canada, 2022.

\bibitem[\protect\citeauthoryear{Foscarin}{2020}]{foscarin2020musical}
Francesco Foscarin.
\newblock {\em The Musical Score: a challenging goal for automatic music
  transcription}.
\newblock PhD thesis, Paris, CNAM, 2020.

\bibitem[\protect\citeauthoryear{Gray and Bunescu}{2016}]{gray2016neural}
Patrick Gray and Razvan~C Bunescu.
\newblock A neural greedy model for voice separation in symbolic music.
\newblock In {\em {Proceedings of the International Society for Music
  Information Retrieval Conference (ISMIR)}}, 2016.

\bibitem[\protect\citeauthoryear{Hamilton \bgroup \em et al.\egroup
  }{2017}]{hamilton2017representation}
William~L. Hamilton, Rex Ying, and Jure Leskovec.
\newblock Representation learning on graphs: Methods and applications.
\newblock {\em IEEE Data Engineering Bulletin}, 40(3):52--74, 2017.

\bibitem[\protect\citeauthoryear{Han \bgroup \em et al.\egroup
  }{2004}]{han2004algorithm}
Mei Han, Wei Xu, Hai Tao, and Yihong Gong.
\newblock An algorithm for multiple object trajectory tracking.
\newblock In {\em Proceedings of the IEEE Computer Society Conference on
  Computer Vision and Pattern Recognition}, volume~1. IEEE, 2004.

\bibitem[\protect\citeauthoryear{Hastie \bgroup \em et al.\egroup
  }{2009}]{hastie2009elements}
Trevor Hastie, Robert Tibshirani, Jerome~H Friedman, and Jerome~H Friedman.
\newblock {\em The elements of statistical learning: data mining, inference,
  and prediction}, volume~2.
\newblock Springer, 2009.

\bibitem[\protect\citeauthoryear{Hsiao and Su}{2021}]{hsiao2021learning}
Yo-Wei Hsiao and Li~Su.
\newblock Learning note-to-note affinity for voice segregation and melody line
  identification of symbolic music data.
\newblock In {\em {Proceedings of the International Society for Music
  Information Retrieval Conference (ISMIR)}}, 2021.

\bibitem[\protect\citeauthoryear{Huron}{2001}]{huron2001tone}
David Huron.
\newblock Tone and voice: A derivation of the rules of voice-leading from
  perceptual principles.
\newblock {\em Music Perception}, 19(1):1--64, 2001.

\bibitem[\protect\citeauthoryear{Jeong \bgroup \em et al.\egroup
  }{2019}]{jeong2019graph}
Dasaem Jeong, Taegyun Kwon, Yoojin Kim, and Juhan Nam.
\newblock Graph neural network for music score data and modeling expressive
  piano performance.
\newblock In {\em Proceedings of the International Conference on Machine
  Learning (ICML)}, 2019.

\bibitem[\protect\citeauthoryear{Jordanous}{2008}]{jordanous2008voice}
Anna Jordanous.
\newblock Voice separation in polyphonic music: A data-driven approach.
\newblock In {\em Proceedings of the International Computer Music Conference
  (ICMC)}, 2008.

\bibitem[\protect\citeauthoryear{Karystinaios and Widmer}{2022}]{cadence2022}
Emmanouil Karystinaios and Gerhard Widmer.
\newblock Cadence detection in symbolic classical music using graph neural
  networks.
\newblock In {\em {Proceedings of the International Society for Music
  Information Retrieval Conference (ISMIR)}}, 2022.

\bibitem[\protect\citeauthoryear{Kilian and Hoos}{2002}]{kilian2002voice}
J{\"u}rgen Kilian and Holger~H Hoos.
\newblock Voice separation-a local optimization approach.
\newblock In {\em {Proceedings of the International Society for Music
  Information Retrieval Conference (ISMIR)}}. Citeseer, 2002.

\bibitem[\protect\citeauthoryear{Liu \bgroup \em et al.\egroup
  }{2022}]{liu2022glan}
He~Liu, Tao Wang, Congyan Lang, Songhe Feng, Yi~Jin, and Yidong Li.
\newblock Glan: A graph-based linear assignment network.
\newblock {\em arXiv preprint arXiv:2201.02057}, 2022.

\bibitem[\protect\citeauthoryear{Ma \bgroup \em et al.\egroup
  }{2022}]{ma2022robust}
Xichu Ma, Xiao Liu, Bowen Zhang, and Ye~Wang.
\newblock Robust melody track identification in symbolic music.
\newblock In {\em {Proceedings of the International Society for Music
  Information Retrieval Conference (ISMIR)}}, 2022.

\bibitem[\protect\citeauthoryear{Madsen and
  Widmer}{2006}]{madsen2006separating}
S{\o}ren~Tjagvad Madsen and Gerhard Widmer.
\newblock Separating voices in midi.
\newblock In {\em {Proceedings of the International Society for Music
  Information Retrieval Conference (ISMIR)}}. Citeseer, 2006.

\bibitem[\protect\citeauthoryear{McLeod and Steedman}{2016}]{mcleod2016hmm}
Andrew McLeod and Mark Steedman.
\newblock Hmm-based voice separation of midi performance.
\newblock {\em Journal of New Music Research}, 45(1):17--26, 2016.

\bibitem[\protect\citeauthoryear{Rafailidis \bgroup \em et al.\egroup
  }{2009}]{rafailidis2009musical}
Dimitris Rafailidis, Emilios Cambouropoulos, and Yannis Manolopoulos.
\newblock Musical voice integration/segregation: Visa revisited.
\newblock In {\em Proceedings of the Sound and Music Computing Conference
  (SMC)}, 2009.

\bibitem[\protect\citeauthoryear{Schlichtkrull \bgroup \em et al.\egroup
  }{2018}]{schlichtkrull2018modeling}
Michael Schlichtkrull, Thomas~N Kipf, Peter Bloem, Rianne Van Den~Berg, Ivan
  Titov, and Max Welling.
\newblock Modeling relational data with graph convolutional networks.
\newblock In {\em Proceedings of the European Semantic Web Conference}.
  Springer, 2018.

\bibitem[\protect\citeauthoryear{Shooner \bgroup \em et al.\egroup
  }{2010}]{shooner2010high}
Christopher Shooner, Srimant~P Tripathy, Harold~E Bedell, and Haluk
  {\"O}{\u{g}}men.
\newblock High-capacity, transient retention of direction-of-motion information
  for multiple moving objects.
\newblock {\em Journal of Vision}, 10(6):1--20, 2010.

\bibitem[\protect\citeauthoryear{Temperley}{2008}]{temperley2008probabilistic}
David Temperley.
\newblock A probabilistic model of melody perception.
\newblock {\em Cognitive Science}, 32(2):418--444, 2008.

\bibitem[\protect\citeauthoryear{Van~der Merwe and
  De~Villiers}{2016}]{van2016comparative}
Lodewyk Van~der Merwe and Pieter De~Villiers.
\newblock Comparative investigation into viterbi based and multiple hypothesis
  based track stitching.
\newblock {\em IET Radar, Sonar \& Navigation}, 10(9):1575--1582, 2016.

\bibitem[\protect\citeauthoryear{Wang \bgroup \em et al.\egroup
  }{2021}]{wang2021joint}
Yongxin Wang, Kris Kitani, and Xinshuo Weng.
\newblock Joint object detection and multi-object tracking with graph neural
  networks.
\newblock In {\em Proceedings of the IEEE International Conference on Robotics
  and Automation (ICRA)}. IEEE, 2021.

\bibitem[\protect\citeauthoryear{Weng \bgroup \em et al.\egroup
  }{2021}]{weng2021ptp}
Xinshuo Weng, Ye~Yuan, and Kris Kitani.
\newblock Ptp: Parallelized tracking and prediction with graph neural networks
  and diversity sampling.
\newblock {\em IEEE Robotics and Automation Letters}, 6(3):4640--4647, 2021.

\bibitem[\protect\citeauthoryear{Xu \bgroup \em et al.\egroup
  }{2018}]{xu2018jumping}
Keyulu Xu, Chengtao Li, Yonglong Tian, Tomohiro Sonobe, Ken-ichi Kawarabayashi,
  and Stefanie Jegelka.
\newblock Representation learning on graphs with jumping knowledge networks.
\newblock In {\em Proceedings of the International Conference on Machine
  Learning (ICML)}, 2018.

\end{thebibliography}

\end{document}